\documentclass[letterpaper, 10pt, conference]{ieeeconf}
\IEEEoverridecommandlockouts 
\usepackage{amsmath,amscd,amssymb,amsgen,amsfonts,amsbsy}
\usepackage{mathrsfs}
\usepackage[utf8x]{inputenc}
\usepackage{color}
\usepackage{textcomp}
\usepackage{float}
\usepackage{latexsym,graphicx}

\usepackage{tikz}
\usetikzlibrary{automata,arrows,positioning,calc}
\usepackage{url}
\usepackage{cite}
\usepackage{algorithmic}
\usepackage[ruled,lined]{algorithm2e}

\newcommand{\remove}[1]{}

\newcommand{\comments}[1]{}

\usetikzlibrary{shapes.geometric}

\tikzset{
	buffer/.style={
		draw,
		shape border rotate=120,
		isosceles triangle,
		isosceles triangle apex angle=69,
		fill=white,
		node distance=10cm,
		minimum height=10em
	}
}

\makeatother

\title{Monte Carlo Rollout Policy for Recommendation Systems with Dynamic User Behavior}
\author{Rahul Meshram and Kesav Kaza
\thanks{Email: rahulmeshram07@gmail.com, kesav.kaza@gmail.com  
}
 }

\begin{document}

\maketitle

\begin{abstract}
 We model online recommendation systems using the hidden Markov multi-state restless multi-armed bandit problem. To solve this we present Monte Carlo rollout policy.  We illustrate numerically that Monte Carlo rollout policy performs better than myopic policy for arbitrary transition dynamics with no specific structure. But, when some structure is imposed on the  transition dynamics,   myopic policy performs better than Monte Carlo rollout policy.
   
\end{abstract}

\section{Introduction}
Online recommendation systems (RS) are extensively used by multimedia hosting platforms e.g. YouTube, Spotify,  and entertainment services e.g. Netflix, Amazon Prime etc.  These systems create personalized playlists for users based on user behavioral information from individual watch history and also by harvesting information from social networking sites. In this paper we provide new models for user behavior and algorithms for recommendation.

Most often  playlists are generated using the ``matrix completion'' problem and items are recommended to users   based on their past preferences. It is implicitly assumed that user interest is static and  the current recommendation does not influence future behavior of user interest. So, a playlist that is generated does not take into account the dynamic behavior or changes in user interest triggered by the current recommendation. In this paper we study a playlist generation system  as a recommendation system, where a playlist is generated using immediate dynamic behavior of user interest. The user responds to different items differently. This behavior depends on the play history along with some element of randomness in the preferences. 

We consider a Markov model for user interest or preferences where a state describes the intensity level of preferences\footnote{ Markov model is an approximation of the dynamic behavior of user interest to make the analysis tractable. In general, user behavior can be more complex and requires further investigation.}. A higher state means higher level of interest for an item. The user behavior for an item is determined by the  transition dynamics for that item. We assume that the user provides a binary feedback upon the play of an item,  and no feedback from not playing it. \footnote{In general there can be other forms of feedback such as the user stopping a video in between, etc.} 

The likelihood of observing feedback is state dependent. The user interest goes to different states with different probability after playing an item. User interest for an item returns to a fixed state whenever it is not played. The item for which user interest stays in higher state with high probability after its play, is  referred as a \textit{viral item}. For certain items, user interest drops immediately after playing it; these are referred as \textit{normal items}.  Our objective is to model and analyze the diverse behavior of user interest for different items, and generate a dynamic playlist using binary feedback. Note that the state of user interest is not observable by the recommendation system. This is an example of multi-state hidden Markov model.  Our model here is a generalization of the two-state hidden Markov model in \cite{Meshram17a}. This paper studies a playlist generation system using multi-state hidden Markov model. 


 
We make following contributions in this paper. 
\begin{enumerate}
	\item We model a playlist generation (recommendation) system as hidden Markov multi-state restless multi armed bandit problem. We present a four state model. An item in an RS is modeled using a POMDP. This is given in Section~\ref{sec:Reco-RMAB}. 
	
	\item  We present the following solution approaches---myopic policy,  Monte Carlo rollout policy and Whittle-index policy in Section~\ref{sec:Solution}. The Whittle-index policy has limited applicability due to lack of an explicit index formula for multi-state hidden Markov bandits.
	
	\item We discuss numerical examples in Section~\ref{sec:Numerical}. 
	We present numerical results with myopic and Monte Carlo rollout policy. 
	Our first numerical example illustrates that myopic policy performs better than Monte Carlo rollout policy whenever transition probabilities of interest states have a specific  structure such as stochastic dominance.  But myopic policy performs poorly compared to Monte Carlo rollout policy whenever there is no such structure imposed on the model. This is demonstrated in the second numerical example. 
		In the third example, we compare Monte Carlo rollout policy with Whittle index policy and we observe that Monte Carlo policy performs better than Whittle index policy and this is due approximations involved in  index calculations.
	
\end{enumerate}


\subsection{Related Work}
Recommendation systems are often studied using collaborative filtering methods, \cite{Aggarwal16,Konstan97,Sarwar01}. Matrix factorization (MF) is one such method employed in collaborative filtering, \cite{Coren09,Schafer07}. The idea is to represent a matrix as users and items. Each entry there describes the user rating for an item. 
MF method then transforms a large dimensional matrix into a lower dimensional matrix. 
Machine learning techniques are used in  MF and collaborative filtering, \cite{He17}. Recommendation systems ideas  inspired on work of Matrix completion problem \cite{Candes10}. In all of these models are based on data which is obtained from previous recommendations or historical data. These works assume that user preferences are static and it does not take into account the dynamic behavior of user based on feedback from preceding recommendations. 

Recently, there is another body of work on modeling online recommendation systems. This work is inspired from online learning with bandit algorithms, \cite{Langford07,Li10,Glowacka19}. It uses contextual epsilon greedy algorithms for news recommendation.
Another way to model online recommendation systems such as playlist generation systems are restless multi-armed bandits, \cite{Meshram17a}. In all these systems, user interest dynamically evolves and this evolution is dependent on whether an item is recommended or not.  

We now describe some related work on RMAB, hidden Markov RMAB and their solution methodologies. RMAB is extensively studied for various application  of communication systems, queuing networks and resource allocation problems, \cite{Whittle88,Gittins11}. RMAB problem is NP-hard, \cite{Papadimitriou99}, but heuristic index based policy is used. To use such index based policy, there is requirement of structure on the dynamics of restless bandits. This can be limitation for hidden Markov RMAB, when each restless bandit is modeled using POMDP and  it is very difficult to obtain structural results. This motivates us to look for an alternative policy, and Monte Carlo rollout policy is studied in this work. Monte Carlo rollout policy has been developed for complex Markov decision processes in \cite{Tesauro96,Chang04,Bertsekas20b,Meshram2020}.


\section{Online Recommendation System  as Restless Multi-armed Bandits} 
\label{sec:Reco-RMAB}
We present models of online recommendation systems (RS). There are different types of items to be recommended.  A model for each type describes the specific user behavior for that type of item. 
We consider a four state model where a state  represents the user interest for an item. The states are called as Low interest (L), Medium interest (M), High interest (H) and Very high interest (V). Thus, the state space is $S = \{L,M, H,V\}.$ RS can play an item or not play that item. The state evolution of user interest for an item depends on actions. There are two actions for each item, play or not play, i.e., $A = \{0,1\},$ where $0$ corresponds to not playing and $1$ corresponds to play of item. 


We suppose that RS gets a binary observation signal, i.e., $1$ for like and $0$ for dislike\footnote{These observations dictate the actions of user based on their interest. For example, the user may skip an item when he dislikes it, or watch completely when he likes it. Further, more signals correspond to more actions from user.}. In general, RS can have more than two signals as observations but for simplicity we consider only two signals. RS can not directly observe user interest for items and hence the state of each item is not observable. When an item is played the user clicks on either like or dislike (skip) buttons with probability $\rho_{i},$ and this click-through probability depends on current state of user interest $i$  for that item but not on the state of any other of items. Whenever user clicks, RS accrues a unit reward with probability $\rho_i$ for $i \in S.$ Further, we assume $\rho_L < \rho_M < \rho_H< \rho_V.$ Thus, each item can be modeled as partially observable Markov decision process (POMDP) and it has finite states with two actions. From literature on POMDP \cite{Smallwood-Sondik73,Sondik78,Lovejoy87}, a belief vector $\pi = (\pi(1),\pi(2), \pi(3), \pi(4))$ is maintained for each item, where $\pi(i)$ is the probability that user interest for the item is in state $i$ and $\sum_{i\in S}\pi(i) = 1.$ The immediate expected reward to RS from play of an item with belief $\pi$  is $\rho(\pi) = \rho_L \pi(L) + \rho_M \pi(M) +  \rho_H \pi(L) +\rho_V \pi(V).$ 

When an item is not played, this implies that another item is played to the user. In this way the items  are competing at RS for each time slot. The user interest state evolution of each item is dependent on whether that item is played or not.   RS is an example of \textit{restless multi-armed bandit problem (RMAB)}, \cite{Whittle88}.

Suppose there are $N$ independent items, each item has the same number of states. 
After each play of an item, a unit reward is obtained by RS based on the user click. Further, RS can play only one arm (item) at each time instant. The objective of RS is to maximize the long term discounted cumulative reward (sum of cumulative reward from play of all items over the long term) subject to constraint that only one item is played at a time. Because RS does not observe the state of each user interest for item at each time step, we refer to this as \textit{hidden Markov RMAB}, \cite{Meshram18}.  This is a constrained optimization problem and the items are coupled due to the  integer constraint on RS. It is also called as \textit{weakly coupled POMDPs.}

\section{Solution Approach} 
\label{sec:Solution}
We  discuss the following solution approaches---myopic policy, Monte-Carlo rollout policy and Whittle index policy.  

We first describe the belief update rule for an  item. 
After play of an item, the belief is 
%
$	\pi_{t+1}(l) 
	= \frac{\sum_{i\in S} \pi_t(i)  p_{i,l}^1 \rho_i}{\sum_{j \in S} \sum_{i\in S} \pi_t(i)  p_{i,j}^1 \rho_i},$
%
and  $\pi_{t+1} =(\pi_{t+1}(L), \pi_{t+1}(M),\pi_{t+1}(H),\pi_{t+1}(V)).$ Here, $\pi_t$ is the belief vector at time $t$ and $P^1=[[p_{i,j}^1]]$ is the transition probability matrix for an item.  
For not playing, no signal is observed and hence 
the posterior belief $\pi_{t+1} = \pi_t P^0.$

\subsection{Myopic policy} 
This is the simplest policy for RMAB with hidden states. In any given slot the item with the highest immediate expected payoff is played.
Let $\pi_{j,t}$  be the belief vector for item $j$ at time $t.$ A unit reward is obtained from playing item $j$ depending on state, with prob. $\rho_j = [\rho_{L,j}, \rho_{M,j}, \rho_{H,j}, \rho_{V,j}].$ Thus, the immediate expected payoff from play of item $j$ is  $\sum_{i \in S}\pi_{j,t}(i) \rho_{i,j}.$  The myopic policy plays an item 
\begin{eqnarray*}
	j^* = \arg\max_{ 1 \leq j \leq N}  \sum_{i \in \mathcal{S}}\pi_{j,t}(i) \rho_{i,j}.
\end{eqnarray*}

\subsection{Look ahead Policy using Monte Carlo Method} 

We study Monte Carlo rollout policy. There are $L$ trajectories simulated for fixed horizon length $H,$ using a known transition and reward model. Along each trajectory, a fixed policy $\phi$ is employed according to which one item is played at each time step.  
The  information obtained from a single trajectory upto horizon length $H$ is
\begin{eqnarray} 
 \{ \pi_{j,t,l},a_{j,t,l}, R_{j,t,l}^{\phi}\}_{j=1,t=1}^{ N,  H}
\end{eqnarray} 
  under policy $\phi.$ Here, $l$ denotes a trajectory. The value estimate of  trajectory $l$ starting from belief state $\pi = (\pi_{1}, \cdots, \pi_N),$ for $N$ items and initial action $a \in \{1,2, \cdots,N \}$ is   
\begin{eqnarray*}
	Q_{H,l}^{\phi}(\pi, a) &=& \sum_{h=1}^{H} \beta^{h-1} R_{h,l}^{\phi} 
	= \sum_{h=1}^{H} \beta^{h-1} r(\pi_{h,l},a_{h,l}, \phi).
\end{eqnarray*}  
Then, the value estimate for state $\pi$ and action $a$ over $L$ trajectories under policy $\phi$ is 
\begin{eqnarray*}
	\widetilde{Q}_{H,L}^{\phi}(\pi, a) = \frac{1}{L}\sum_{l=1}^{L}  Q_{H,l}^{\phi}(\pi, a, W).
\end{eqnarray*} 
Here, policy $\phi$ can be uniform random policy or myopic (greedy) policy that is implemented for a trajectory.  Next, a one step policy improvement is performed, and the optimal action selected is according follow rule. 

\begin{eqnarray}
j^*(\pi) = \arg \max_{1 \leq j \leq N} \left[ r(\pi, a= j) + \beta \widetilde{Q}_{H,L}^{\phi}(\pi, a = j) \right].
\end{eqnarray}

In each time step, an item is  played based on the above rule. 
Detailed discussion on rollout policy for RMAB is given in \cite{Meshram2020}.

\subsection{Whittle-index policy}

Another popular approach for RMAB (and weakly coupled POMDPs) is Whittle-index policy \cite{Whittle88}, where the constrained optimization problem can be solved via relaxing the integer constraints. The problem is transformed into an optimization problem with discounted constraints. Later, using Lagrangian technique, one decouples relaxed constrained optimization RMAB problem into $N$ single-armed restless bandit problems.  In a single-armed restless bandit (SARB) problem, a subsidy $W$  for not playing the item is introduced. A SARB with hidden states is an example of a POMDP with a two action model, \cite{Meshram18}. To use Whittle index policy, one requires to study structural properties of SARB,  show the existence of a threshold policy,  indexability for each item, and   compute the indices for all items. In each time step, the item with highest index is played. 

In our model, it is very difficult to claim indexability and obtain closed form index formula. The idexability condition require us to show a threshold policy behavior for each item. In \cite[Proposition $1$ and Lemma $2.1$]{Lovejoy87}, authors have shown existence of threshold policy for specialized model in POMDP.    
In a specialized model, it is possible to show indexability (detail is omitted) and use Monte Carlo based index computation algorithm,  see \cite[Section IV, Algorithm $1$]{Meshram2020}. Note that this algorithm is computationally expensive and time consuming  because Monte-Carlo algorithm has to run for each restless bandits till their value function converges.

\section{Numerical Results for Model $1$}
\label{sec:Numerical}

We now present numerical examples that illustrate the performance of myopic policy and Monte-Carlo rollout policy.  In the first example we observe that myopic policy performs better than MC rollout policy for some structural assumptions on transition probabilities and reward probabilities. Finally, in the second example there are no structural assumptions on the transition probabilities. Here the MC rollout policy performs better than myopic policy.

\subsubsection{Example-$1$}

In this example, we  introduce structure on transition probability matrices of items, as in \cite{Lovejoy87}.
When an item is played, the user interest evolves according to different transition matrix corresponding to different items. But for not played items, the user interest evolves according to a common transition matrix. We use the following parameter set.  The number of items $N =5,$ number of states $S= 4.$  $\beta = 0.95$
Transition probability for items when that is played is denoted by $P^{1}_j,$ $j=1,2, \cdots,N.$  When item is not played, then the transition probability matrix is $P^0$ and it is same for all items.

{\footnotesize{
\begin{eqnarray*}
	P_1^1 = \left[ \begin{array}{cccc}
		0.3 & 0.7 & 0 & 0  \\
		0.2 & 0.3 & 0.5 & 0 \\
		0 & 0.1 & 0.4 & 0.5 \\
		0 & 0 & 0.25 & 0.75 
	\end{array} \right] ,
%
	P_2^1 = \left[ \begin{array}{cccc}
		0.1 & 0.9 & 0 & 0  \\
		0.3 & 0.35	 & 0.35 & 0 \\
		0 & 0.25 & 0.25 & 0.5 \\
		0 & 0 & 0.25 & 0.75 
	\end{array} \right] ,
\end{eqnarray*}
}}
{\footnotesize{
\begin{eqnarray*}
	P_3^1 = \left[ \begin{array}{cccc}
		0.45 & 0.55 & 0 & 0  \\
		0.3 & 0.3 & 0.4 & 0 \\
		0 & 0.2 & 0.35 & 0.45 \\
		0 & 0 & 0.1 & 0.9 
	\end{array} \right] 
%
	P_4^1 = \left[ \begin{array}{cccc}
		0.5 & 0.5 & 0 & 0  \\
		0.1 & 0.4 & 0.5 & 0 \\
		0 & 0.3 & 0..3 & 0.4 \\
		0 & 0 & 0.4 & 0.6 
	\end{array} \right] 
\end{eqnarray*}
}}
{\footnotesize{
\begin{eqnarray*}
	P_5^1 = \left[ \begin{array}{cccc}
		0.4 & 0.6 & 0 & 0  \\
		0.25 & 0.35 & 0.4 & 0 \\
		0 & 0.3 & 0.35 & 0.35 \\
		0 & 0 & 0.45 & 0.55 
	\end{array} \right] 
	P^0 = \left[ \begin{array}{cccc}
		0.45 & 0.55 & 0 & 0  \\
		0.15 & 0.4 & 0.45 & 0 \\
		0 & 0.2 & 0.3 & 0.5 \\
		0 & 0 & 0.4 & 0.6 
	\end{array} \right] .
\end{eqnarray*}
}}

Reward vector for all items. 
\begin{eqnarray*}
	\rho = \left[ \begin{array}{cccc}
		0.1 & 0.3 & 0.6 & 0.85 \\
		0.25 & 0.5 & 0.5 & 0.7 \\
		0.2 & 0.6 & 0.6 & 0.6 \\
		0.3 & 0.35 & 0.55 & 0.65 \\
		0.25 & 0.4 & 0.6 & 0.95 
	\end{array}
	\right].
\end{eqnarray*}
Initial belief 
\begin{eqnarray*}
	\pi = \left[ \begin{array}{cccc}
		0.1 & 0.2  & 0.3 & 0.4 \\
		0.3 &  0.25 & 0.4 & 0.05 \\
		0.15 & 0.1 & 0.3 & 0.45\\
		0.5 & 0.1 & 0.1 & 0.3 \\
		0.25 & 0.25 & 0.25 & 0.25 
	\end{array} \right]  
\end{eqnarray*}

Initial state of items from different states. 
$	X = \left[  2, 1, 3, 2,1 \right].$ 
%
%
\begin{figure}
	\begin{center}
			\includegraphics[width=0.9\columnwidth]{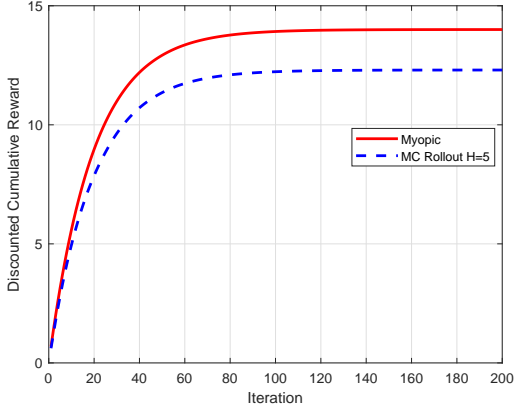}
	\end{center}
	\caption{Comparison of myopic policy and Monte Carlo rollout policy for Example $1$ with $H=5.$ }
	\label{plots:Myopic-MC-Ex1}
\end{figure}
From~Fig.~\ref{plots:Myopic-MC-Ex1}, we find that the myopic policy performs better than Monte Carlo rollout policy. In Monte Carlo rollout policy, we used $H =5$ and $L=100.$


\subsubsection{Example-$2$}
We consider a general transition probability matrix for each action, and with no structural assumption. Hence, we do not have stochastic dominance condition for the transition probability matrix of each item. When item is played, the user interest evolves according to different transition matrices for different items  but for not played items, the user interest evolves according to a common transition matrix. We use the following parameter set.  The number of items $N =5,$ number of states $S= 4.$ We use following parameters. $\beta = 0.95$
Transition probability for items when that is played is denoted by $P^{1}_j,$ $j=1,2, \cdots,N.$  When item is not played, then the transition probability matrix is $P^0$ and it is same for all items. 

{\footnotesize{
\begin{eqnarray*}
	P_1^1 = \left[ \begin{array}{cccc}
		0.7 & 0.3 & 0 & 0  \\
		0 & 0.7 & 0.3 & 0 \\
		0 & 0 & 0.7 & 0.3 \\
		0 & 0.3 & 0 & 0.7 
	\end{array} \right] ,
%
	P_2^1 = \left[ \begin{array}{cccc}
		0.9 & 0.1 & 0 & 0  \\
		0 & 0.9 & 0.1 & 0 \\
		0 & 0 & 0.9 & 0.1 \\
		0.45 & 0 & 0.45 & 0.1 
	\end{array} \right] ,
\end{eqnarray*}
}}

{\footnotesize{
\begin{eqnarray*}
	P_3^1 = \left[ \begin{array}{cccc}
		0.45 & 0.55 & 0 & 0  \\
		0.3 & 0.3 & 0.4 & 0 \\
		0 & 0.2 & 0.35 & 0.45 \\
		0.9 & 0 & 0 & 0.1 
	\end{array} \right] 
%
	P_4^1 = \left[ \begin{array}{cccc}
		0.5 & 0.5 & 0 & 0  \\
		0.1 & 0.4 & 0.5 & 0 \\
		0 & 0.3 & 0..3 & 0.4 \\
		0.4 & 0 & 0.4 & 0.2 
	\end{array} \right] 
\end{eqnarray*}
}}
{\footnotesize{
\begin{eqnarray*}
	P_5^1 = \left[ \begin{array}{cccc}
		0.4 & 0.6 & 0 & 0  \\
		0.25 & 0.35 & 0.4 & 0 \\
		0 & 0.3 & 0.35 & 0.35 \\
		0 & 0.6 & 0.25 & 0.15 
	\end{array} \right] 
%
	P^0 = \left[ \begin{array}{cccc}
		0.5 & 0.5 & 0 & 0  \\
		0.25 & 0.75 & 0 & 0 \\
		0.2 & 0.8 & 0 & 0 \\
		0.05 & 0.95 & 0 & 0 
	\end{array} \right] 
\end{eqnarray*}
}}

Reward vector for all items. 
\begin{eqnarray*}
\rho = \left[ \begin{array}{cccc}
0.1 & 0.1 & 0.1 & 0.85 \\
0.2 & 0.2 & 0.2 & 0.7 \\
0.3 & 0.3 & 0.6 & 0.6 \\
0.3 & 0.35 & 0.55 & 0.65 \\
0.25 & 0.4 & 0.5 & 0.6 
\end{array}
\right]
\end{eqnarray*}

\begin{figure}
	\begin{center}
			\includegraphics[width=0.9\columnwidth]{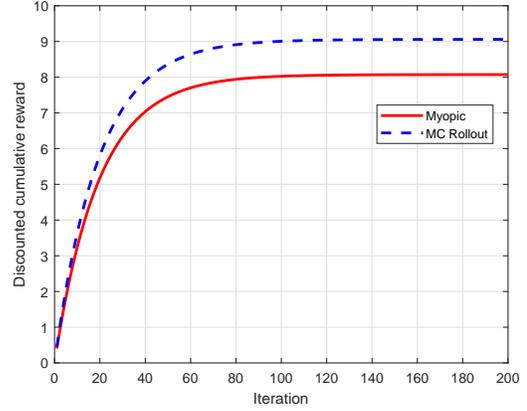}
	\end{center}
	\caption{Comparison of Myopic policy and Monte Carlo rollout policy for Example $2$ with $H=5.$ }
	\label{plots:Myopic-MC-Ex3}
\end{figure}

The initial belief vector and initial state is same as in example-$1.$ 
%
We compare expected discounted cumulative reward with myopic and Monte Carlo rollout policy in Fig.~\ref{plots:Myopic-MC-Ex3}. For Monte-Carlo rollout policy we use length of a trajectory $H =5$ and number of trajectories $L =100.$
With myopic policy, items $3$ and $5$ are played most frequently, whereas with  MC rollout policy, item $4$ is played most frequently. 

\subsubsection{Example $3$}
In this example we use same transition probability matrix as in example-1 when item is played. But when item is not played, the transition happens to state $2.$ This is different from example $1$ and also reward matrix is different. We use same initial belief and  initial state as in example $1.$ We use discount parameter $\beta =0.95.$  

{\footnotesize{
		\begin{eqnarray*}
			P_1^1 = \left[ \begin{array}{cccc}
				0.3 & 0.7 & 0 & 0  \\
				0.2 & 0.3 & 0.5 & 0 \\
				0 & 0.1 & 0.4 & 0.5 \\
				0 & 0 & 0.25 & 0.75 
			\end{array} \right] ,
			%
			P_2^1 = \left[ \begin{array}{cccc}
				0.1 & 0.9 & 0 & 0  \\
				0.3 & 0.35	 & 0.35 & 0 \\
				0 & 0.25 & 0.25 & 0.5 \\
				0 & 0 & 0.25 & 0.75 
			\end{array} \right] ,
		\end{eqnarray*}
}}
{\footnotesize{
		\begin{eqnarray*}
			P_3^1 = \left[ \begin{array}{cccc}
				0.45 & 0.55 & 0 & 0  \\
				0.3 & 0.3 & 0.4 & 0 \\
				0 & 0.2 & 0.35 & 0.45 \\
				0 & 0 & 0.1 & 0.9 
			\end{array} \right] 
			%
			P_4^1 = \left[ \begin{array}{cccc}
				0.5 & 0.5 & 0 & 0  \\
				0.1 & 0.4 & 0.5 & 0 \\
				0 & 0.3 & 0..3 & 0.4 \\
				0 & 0 & 0.4 & 0.6 
			\end{array} \right] 
		\end{eqnarray*}
}}
{\footnotesize{
		\begin{eqnarray*}
			P_5^1 = \left[ \begin{array}{cccc}
				0.4 & 0.6 & 0 & 0  \\
				0.25 & 0.35 & 0.4 & 0 \\
				0 & 0.3 & 0.35 & 0.35 \\
				0 & 0 & 0.45 & 0.55 
			\end{array} \right] 
			P^0 = \left[ \begin{array}{cccc}
				0 & 1 & 0 & 0  \\
				0 & 1 & 0 & 0 \\
				0 & 1 & 0 & 0 \\
				0 & 1 & 0& 0 
			\end{array} \right] .
		\end{eqnarray*}
}}

Reward vector for all items. 
\begin{eqnarray*}
	\rho = \left[ \begin{array}{cccc}
		0.1 & 0.3 & 0.6 & 0.75 \\
		0.25 & 0.45 & 0.55 & 0.75 \\
		0.2 & 0.6 & 0.6 & 0.7 \\
		0.3 & 0.35 & 0.55 & 0.65 \\
		0.3 & 0.5 & 0.6 & 0.95 
	\end{array}
	\right].
\end{eqnarray*}

\begin{figure}
	\begin{center}
		\includegraphics[width=0.9\columnwidth]{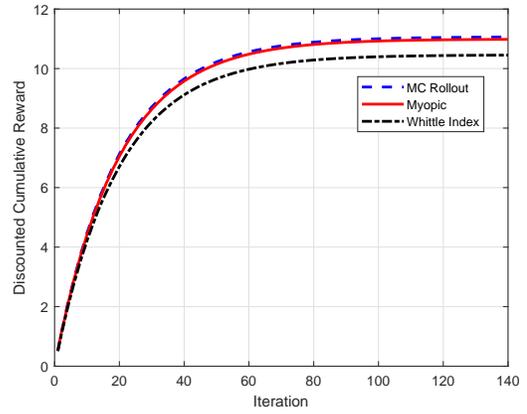}
	\end{center}
	\caption{Comparison of Myopic policy, Monte Carlo rollout policy and Whittle index policy for Example $3$ with $H=5.$ }
	\label{plots:Myopic-MC-Whittle-Ex1}
\end{figure}

We observe from Fig.~\ref{plots:Myopic-MC-Whittle-Ex1} that Myopic policy and Monte Carlo rollout policy performs better than Whittle index policy. This may be due to approximation used for index computation, lack of explicit formula or structure of the problem. As we stated earlier, index policy is more computationally expensive than  Monte Carlo rollout policy and myopic policy when there is no explicit closed formula in case of hidden Markov bandit. In such examples, Monte Carlo rollout policy is good alternative.  The discount parameter $\beta = 0.95$

\subsection{Example $4$}
\begin{figure}
	\begin{center}
		\includegraphics[width=0.9\columnwidth]{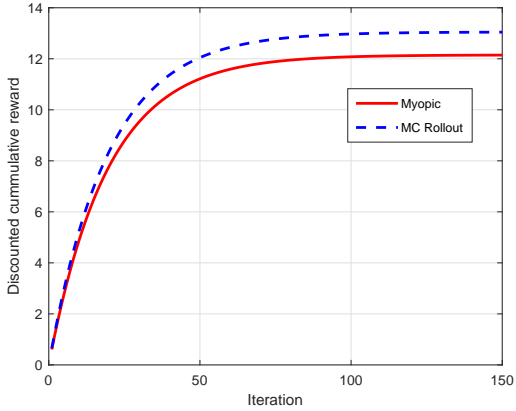}
	\end{center}
	\caption{Comparison of Myopic policy and Monte Carlo rollout policy  for Example $4$ with $H=5.$ }
	\label{plots:Myopic-MC--Ex1-15arms}
\end{figure}

In this example we consider $15$ items with $4$ state. We compare only myopic policy and Monte Carlo rollout policy. We do not assume monotonicity structure on transition matrix. The comparison is illustrated in Fig.~\ref{plots:Myopic-MC--Ex1-15arms}. We observe that Monte-Carlo rollout policy performs  better than myopic policy, i.e, upto $8\%.$ In Monte Carlo rollout policy, we use $H =5$ and $L =100.$ 

\section{Conclusions}

We have studied an online recommendation system problem using hidden Markov RMAB and provided numerical results for Monte Carlo rollout policy and myopic policy. We observed that Monte Carlo rollout policy performs better for arbitrary transition dynamics. We observe numerically that myopic policy performs better than Monte Carlo whenever structure on state transition dynamics. We also presented the performance of Whittle index policy and that is compared with Monte Carlo rollout policy for a specialized model. 


The objective in paper was to describe a new Monte Carlo rollout algorithm for RS with Markov model. We have demonstrated the performance of the algorithm on a small scale example. This study can be extended for large scale examples, e.g., large number of items upto few hundreds. Looking at the scalability problem, even though an RS might have millions of items in its database, it may only recommend items from a small subset considering the cognitive limitations of humans and the problem of information overload.


%
%
%
%


\bibliographystyle{IEEE}

\bibliography{restless-bandits}

%
%
%


\end{document}